\documentclass[12pt]{article}
\usepackage{pic04}
\usepackage{hyperref}
\usepackage{url}
\usepackage{graphicx}

\begin{document}

\title{\bf A METHOD OF EXTRACTING THE MASS OF THE TOP
QUARK IN THE DI-LEPTON CHANNEL USING THE D\O~DETECTOR}
\author{
Sarosh N. Fatakia     \\
{\em Boston University  }      \\
for the D\O~collaboration }   
\maketitle

\baselineskip=14.5pt
\begin{abstract}
We present a method for extracting the mass of the top quark from the 
di-lepton decays of top anti-top quark pairs. In this decay channel 
two neutrinos remain undetected. Extraction of the mass of the top 
quark by kinematic reconstruction is not possible because the event 
is under-constrained. We therefore employ a dynamical likelihood method 
to solve the problem.
\end{abstract}

\baselineskip=17pt

%\newpage

\section{The di-lepton event topology}
In the di-lepton channel top anti-top quark pairs decay via $t\rightarrow Wb,$ 
followed by $W \rightarrow l\nu.$ We can identify and measure the $4$-momenta 
of the jets\footnote{Both jets are assumed to be $b$-jets} and the charged 
leptons. The two neutrinos in the event remain undetected. The vector sum of 
their transverse momenta can be inferred from the observed missing $p_{T}.$
This leaves us with $14$ observables $\{o\}$ out of $18$ values $\{ v \}$ 
needed to describe the six particle final state. In order to constrain the 
$t\bar{t}$ kinematics, three additional constraints on the invariant masses of 
final state particle combinations are introduced: \\
$\star~~m(l^{+}_{1},\nu_{l_{1}}) = m(l^{-}_{2},\bar{\nu}_{l_{2}}) 
 \equiv~ m_{W},$  \\
$\star~~m(b,l^{+}_{1},\nu_{l_{1}}) 
= m(\bar{b}, l^{-}_{2},\bar{\nu}_{l_{2}})$ \\
In the end the set of equations are still under-constrained by one equation. 
Extraction of the mass of the top quark by kinematic reconstruction is 
therefore not possible.

\section{The Analysis}
This analysis was developed during Run I\cite{Run_I}. Ideally we would like 
to calculate the probability to measure the $14$ observables $\{o\}$, given 
the top quark mass $m_{t}.$
\begin{equation}
P(\{o\}| m_{t})  \propto  \int_{ \{ v \} } 
f(x) f(\bar{x}) |{\cal {M}}|^{2} p(\{o\}|\{v\}) 
d^{18}\{v\} \,dx d\bar{x} 
\label{evt_pdf}
\end{equation}
is the probability density function ({\it pdf}) for measuring $\{o\}$ given 
the $18$ final state parameters $\{v\}$ for a hypothetical $m_{t}$ and for 
the parton (anti-parton) momentum fraction $x~(\bar{x}).$ Here $f(x)$ 
$(f(\bar{x}))$ is the proton (anti-proton) parton distribution function and  
${\cal {M}}$ is the matrix element for the process. A hypothetical value of 
$m_{t}$ is used as the last constraint. We then obtain up to $4$ real solutions 
for the $\nu$ and $\bar{\nu}$ momenta\cite{dalitz}. There is a two-fold 
combinatoric ambiguity in pairing the charged leptons and $b$-jets. As a 
result up to eight solutions of the neutrino momenta are possible.
Instead of rigorously computing the {\it pdf} in Equation~\ref{evt_pdf}, 
event weights\cite{kondo}\cite{dalitz} are used to characterize the physics. 
In this poster we calibrate the effect of this simplification by comparing the 
measured mass of the top quark obtained from the likelihood fits to simulated 
events, versus the value used in their generation.

For every event, a weight $({\cal {W}}_{k})$ which is a function of the 
hypothesized mass of the top quark $(m_{t}),$ is 
derived\cite{Run_I}\cite{kondo}\cite{dalitz} that corresponds to the 
$k^{th}$ neutrino momentum solution. The event weight
\begin{equation}
{\cal {W}}_{k}(m_{t}) \propto f(x)f(\bar{x})p(E'| m_t)p(\bar{E}'| m_t),
\end{equation}
represents the likelihood to observe the event for a given $m_{t}.$
Here $p(E'|m_t)$ is the probability density function for the energy of 
the charged lepton to be $E'$ in the rest frame of the top quark of mass 
$m_{t}$\cite{lep_mom}. The quantity $p(\bar{E}'|m_t)$ is the analogue for 
the anti-top quark. With $n$ solutions for the neutrino momenta the total 
event weight ${\mathbf {W}}$ is defined as:
\begin{equation}
{\mathbf {W}} = \mbox{normalization} \sum_{k=1}^{n} ~{\cal {W}}_{k}
\end{equation}

The value of $m_{t}$ at which the weight curve peaks is used as the mass 
estimator in the analysis.
\begin{figure}[htbp]
\begin{center}
  \includegraphics[height=5.0cm ]{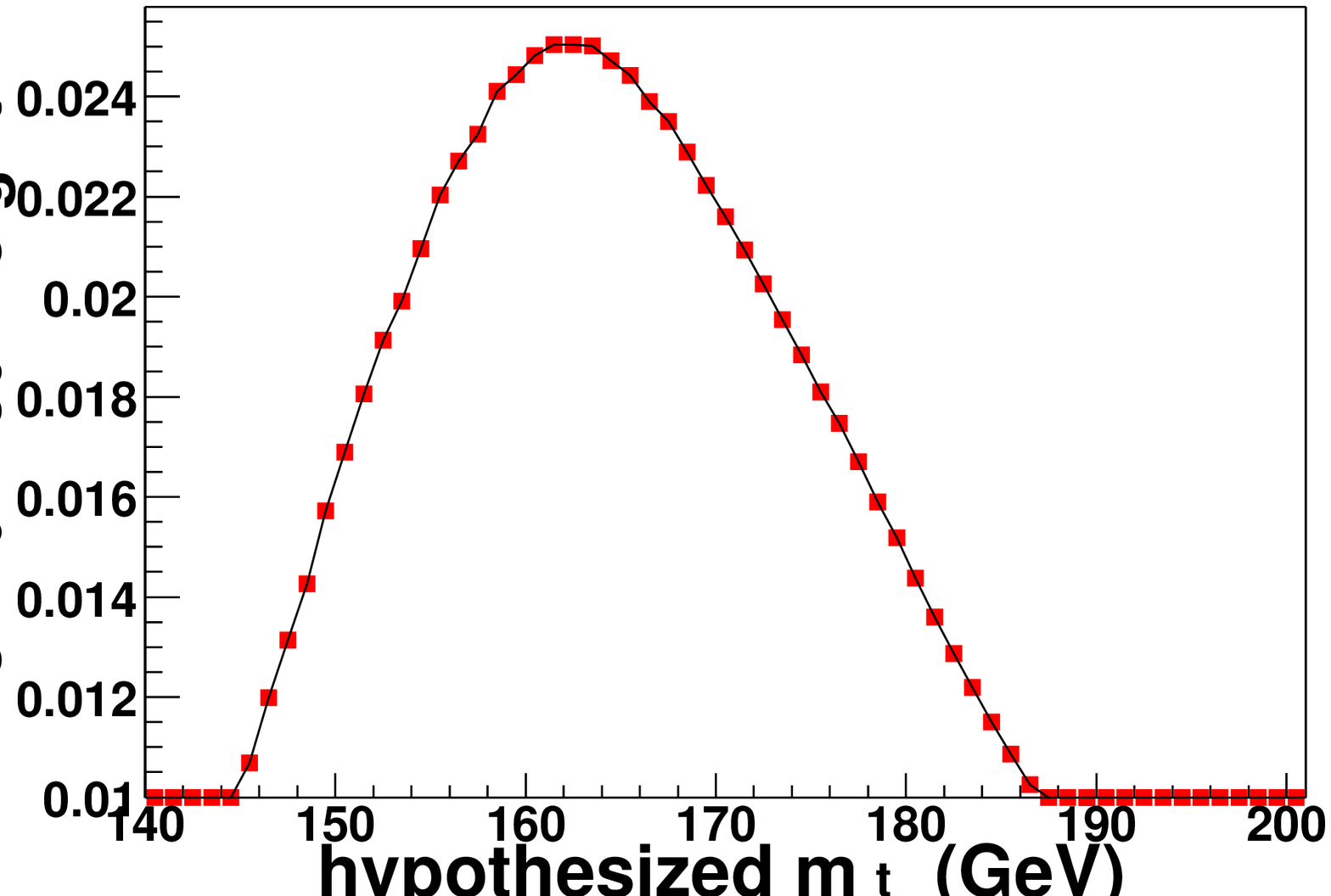}
  \hspace{1.0cm}
  \includegraphics[height=5.0cm ]{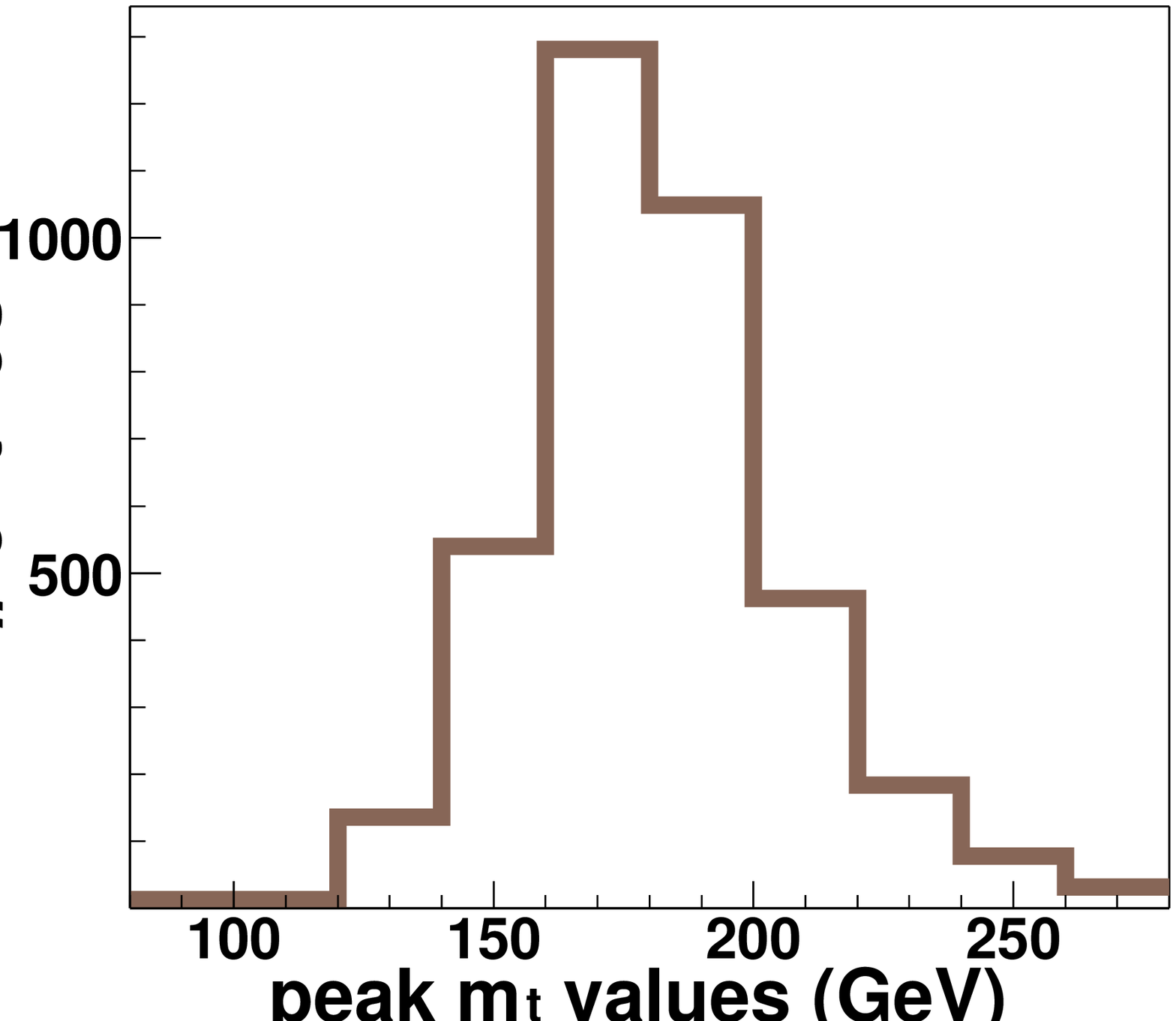}    
  \caption{(left)~Example weight distribution from a simulated event
    generated with $m_{t}=175.0$ GeV.
    (right)~Template constructed from peak values obtained from many events
    each having input $m_{t}=175.0$ GeV.}
\label{peak_plots} 
\end{center}
\end{figure}
All simulated events are first filtered using the same criteria as for 
data events\cite{d0_Xsec_note}. Figure~\ref{peak_plots}(left) shows the 
weight curve for a simulated event with input $m_{t}=$ $175.0$ GeV.
The peak values from many events generated with the same $m_{t}$ are 
binned into histograms to construct templates, one of which is shown in 
Figure~\ref{peak_plots}(right).

It is not possible to generate MC events with continuously
varying input $m_{t}.$ Seven distinct $m_{t}$ values are used 
to generate the signal MC template distributions from $120.0$ 
GeV to $230.0$ GeV.
Templates representing contamination 
from background processes are also constructed and then 
added to the signal templates in the proportion of the
expected background\cite{d0_Xsec_note}. 
Normalized templates represent the likelihood of observing
the particular event if the mass of the top quark equals
$m_{t}.$ 

Sets of eight MC events are used to construct simulated 
experiments\footnote{Eight {\it emu} events observed 
from $142.73~\mbox{pb}^{-1}$ data\cite{d0_Xsec_note}.}. 
\begin{figure}[htbp]
  \centerline{\hbox{ \hspace{0.2cm}
    \includegraphics[height=5.0cm, width=5.0cm ]{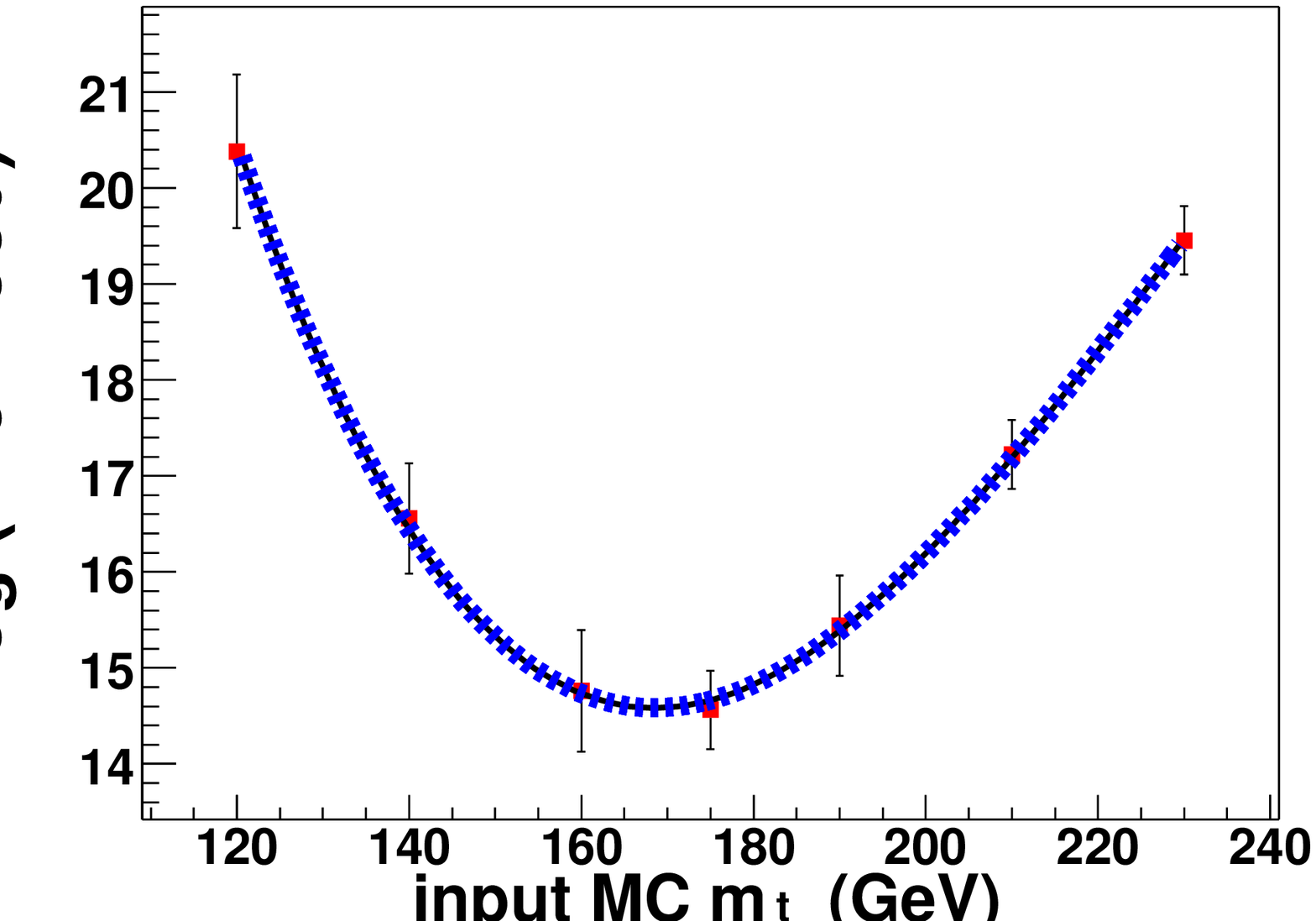}
    \hspace{0.3cm}
    \includegraphics[height=5.0cm, width=4.7cm ]{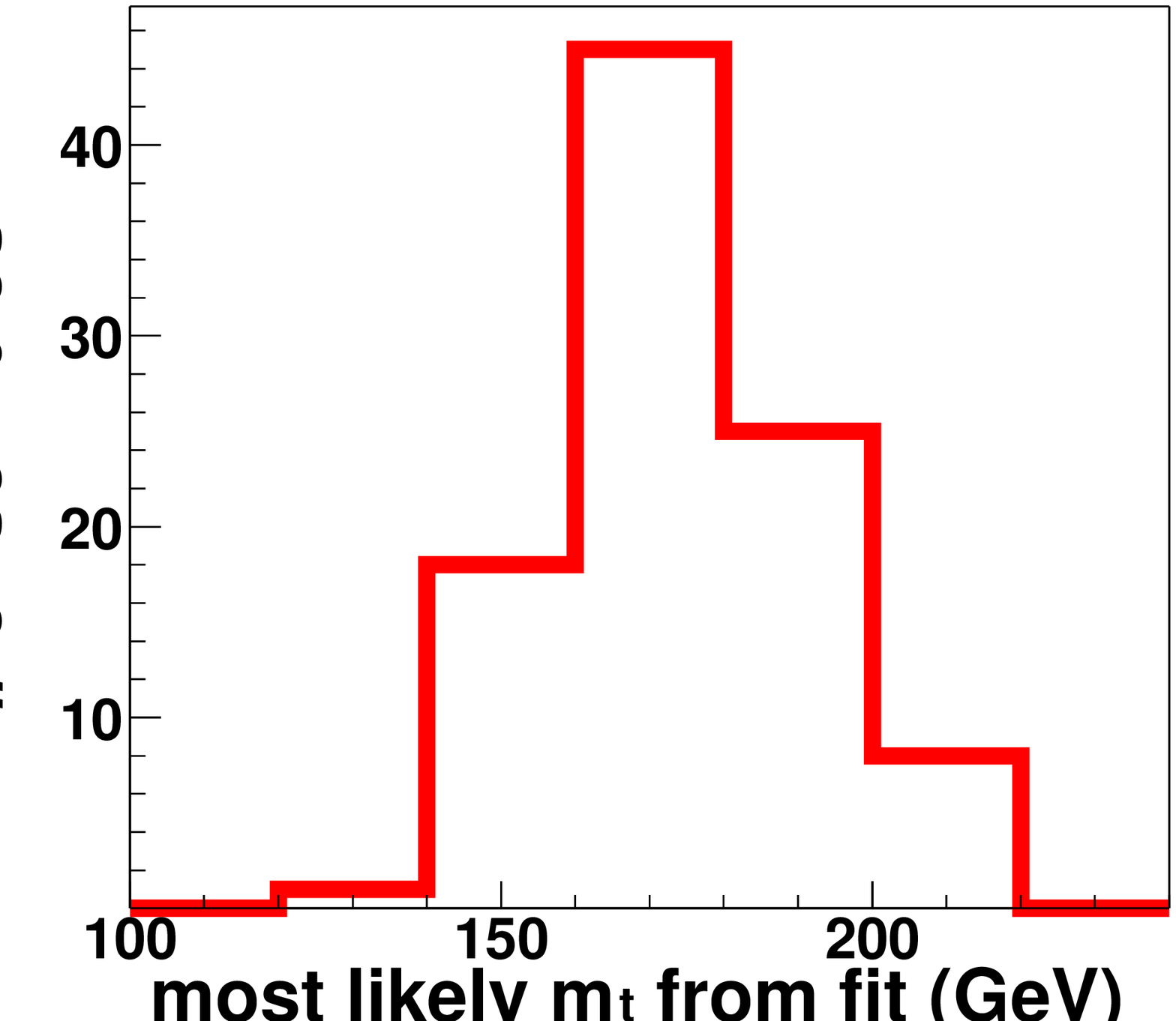}
    \hspace{0.3cm}
    \includegraphics[height=5.0cm, width=5.0cm]{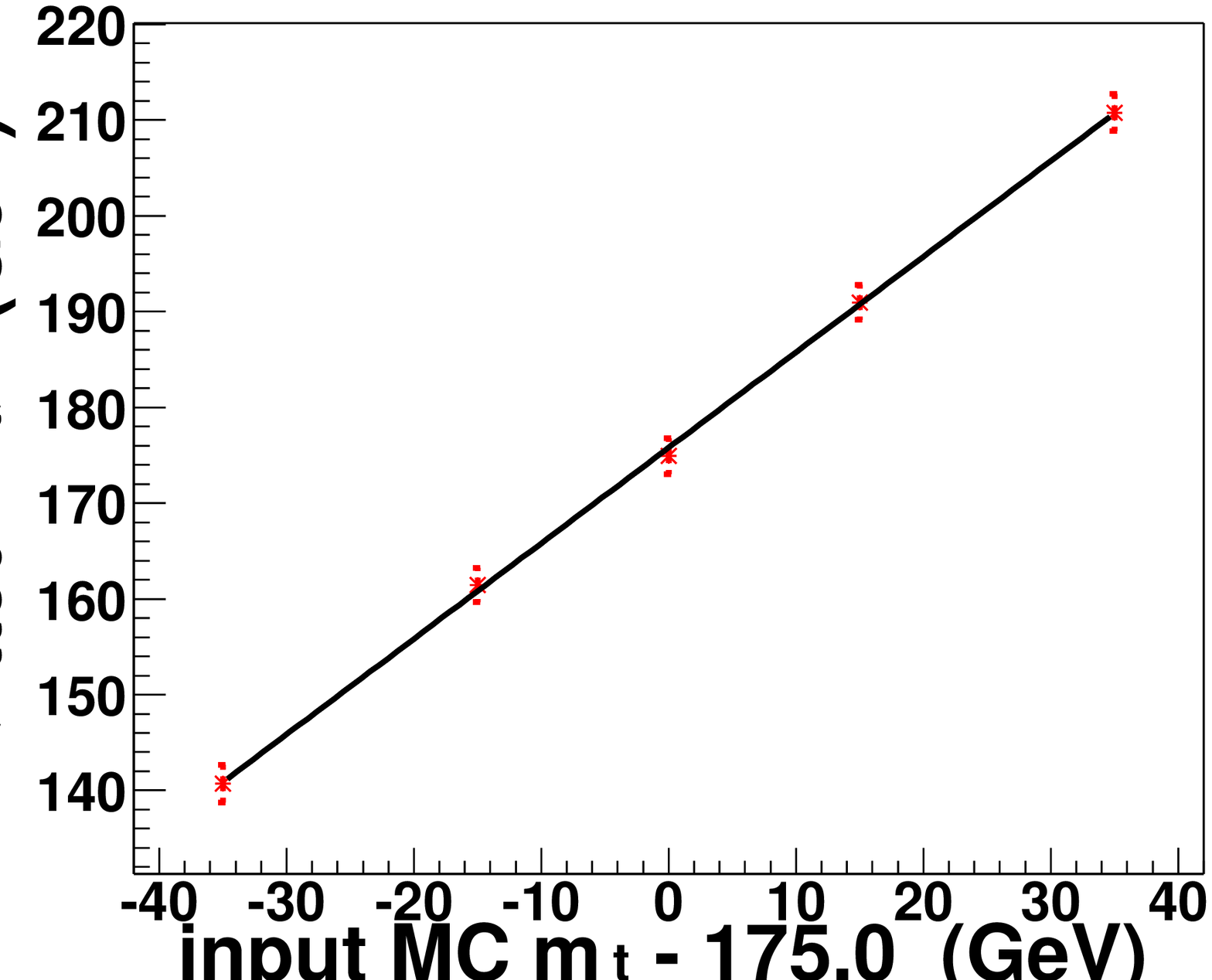}
    }
  }
  \caption{(left) Binned maximum log-likelihood fit of an ensemble of
    events to the template histograms yield the most likely estimate 
    (MLE) of $m_{t}$ for that ensemble. (center) A histogram of MLEs derived
    from $100$ different ensembles, all signal events used have
    an input $m_{t}=175.0$ GeV. (right) Calibration of a series of different
    input $m_{t}$ values. The $175$ GeV input MC value follows from the
    histogram in the center.
  }
  \label{calib_plots} 
\end{figure}
A binned maximum likelihood fit is performed for each such ensemble 
using the template distributions.
One such fit to ensemble events is illustrated in 
Figure~\ref{calib_plots}(left). Many different ensembles are constructed, 
and the exercise 
is repeated many times to obtain a distribution of maximum 
likelihood estimates (MLEs) as in Figure~\ref{calib_plots}(center).
The mean of this distribution yields the mean fitted estimate of $m_{t}$.
Repeating the same exercise for different input MC $m_{t}$ yields 
values which can be used to check the performance of the method.
A straight line parameterized as: \\
{\centerline {
$\mbox{fitted~mass}$
$=p_{1}\cdot(\mbox{input~mass}-175.0~\mbox{GeV})+ p_{0}~\mbox{GeV}$}} \\  
gives the best fit to the ensemble test results 
for $p_{1}=1.00\pm0.04$
and $p_{0}=175.8\pm0.08$ GeV. This fit to the set of points, shown in 
Figure~\ref{calib_plots}(right), is consistent with a straight line 
of unit slope and  $175.0$ GeV offset.
\section{Conclusion}
The calibration of this method verifies its performance, and
indicates absence of bias in the method.
\section{Acknowledgments}
I would like to thank the D\O~collaboration for the 
opportunity to present this work at PIC-2004, and in
particular my thesis adviser, Professor Ulrich Heintz, for 
his guidance.

\end{document}